# Demonstration of the Complementarity of One- and Two-Photon Interference

A. F. Abouraddy, M. B. Nasr, B. E. A. Saleh[†], A. V. Sergienko, and M. C. Teich

*Quantum Imaging Laboratory,[‡] Department of Electrical & Computer Engineering,*

*Boston University,*

*8 Saint Mary's Street, Boston, MA 02215*

The visibilities of second-order (single-photon) and fourth-order (two-photon) interference have been observed in a Young's double-slit experiment using light generated by spontaneous parametric down-conversion and a photon-counting intensified CCD camera. Coherence and entanglement underlie one-and two-photon interference, respectively. As the effective source size is increased, coherence is diminished while entanglement is enhanced, so that the visibility of single-photon interference decreases while that of two-photon interference increases. This is the first experimental demonstration of the complementarity between single- and two-photon interference (coherence and entanglement) in the spatial domain.

[†] Electronic address: besaleh@bu.edu

[‡] URL: http://www.bu.edu/qil

# I. INTRODUCTION

The conventional theory of optical coherence dictates that the visibility of interference fringes equals the degree of coherence [1]. In a Young's double-slit experiment using an incoherent light source, the degree of coherence at the slits is inversely proportional to the angular size of the source. With the recent interest in light sources emitting photon pairs in an entangled state, the issue of the visibility of two-photon interference fringes has come to the fore [2-6]. The visibility of two-photon interference fringes in a Young's double-slit experiment is governed by the degree of entanglement. For light generated by spontaneous parametric down-conversion in a nonlinear crystal, the degree of entanglement is controlled by the size of the source (the width of the pump beam) [5,6]. A smaller source size corresponds to reduced entanglement, and therefore to reduced visibility of two-photon interference.

A basic complementarity between coherence and entanglement underlies the complementarity between single- and two-photon interference [2-6]. This complementarity has its origin in the separability of the coherence function and the two-photon wavefunction. While completely coherent light is characterized by a separable coherence function, a separable two-photon wavefunction corresponds to total lack of entanglement. We have recently shown [6] that the dependence of the second-order coherence function and the two-photon wavefunction on the source size is mathematically identical, so that there is a duality between the two systems. But because of the opposite dependence of coherence and entanglement on separability, the source size plays opposite roles in determining the visibilities of single- and two-photon interference. A source of large area generates light of low coherence, but yields highly entangled photon pairs.

In contrast, a small source emits light of high spatial coherence, but generates poorly entangled photon pairs, since the momentum conservation relations cannot be precise. Interference of light in a two-photon state is generally a mixture of one-photon and two-photon interference, which must be carefully identified, if the complementarity is to be properly assessed. As the degree of entanglement increases, this mixture generates a purely two-photon interferogram, and in the limit of the unentangled state it generates a purely one-photon interferogram. The purpose of this paper is to report the first experimental demonstration of this complementarity and to demonstrate the gradual change from one limit to the other.

Several versions of the Young's double-slit experiment have been recently conducted using entangled photons generated by spontaneous parametric down-conversion. In one configuration [7], one of the beams, say the signal, is transmitted through the slits and coincidence measurements are performed with a moving detector behind the slits and a fixed detector at the idler beam. The visibility of such fringes was found to be dependent on the size of the aperture in the idler beam. The experiment was repeated [8] keeping the detector behind the slits fixed and moving the detector in the idler beam, and the same fringes were observed. This was explained through the concept of a "ghost" source at the location of the fixed detector, either in the signal or idler beams [8, 9]. In another configuration, the slits were placed in the pump beam [10] and it was shown that the coincidence rate at the idler and signal detectors exhibited interference fringes when one was kept fixed and the other scanned. In yet another configuration, the down-conversion was collinear and slits were placed in the paths of both signal and idler beams [11], and a detector recorded the arrival of two photons at the same position. A conceptually similar experiment was also performed in a non-collinear configuration [12] with a pair of double slits, one in the signal beam and the other in the idler beam, and coincidence fringes were observed as the signal detector was scanned with the idler detector fixed.

This paper reports single- and two-photon interference in a Young's double-slit experiment, with the slits placed in both signal and idler beams, and demonstrates the complementarity between entanglement and coherence. The configuration is similar to that in [11] but permits the registration of spatially separated photoevents.

## II. THEORY

Consider the generic double-slit interference setup shown in Fig. 1. Light emitted from a source is directed by a linear optical system of impulse response function $h_1(x_1,x)$ onto two slits at positions $x_1$ and $x_2$. Light transmitted through the slits is directed by a second linear optical system of impulse response function $h_2(x',x)$ onto the detection plane. The overall system has an impulse response function

$$h(x',x) = h_2(x',x_1)h_1(x_1,x) + h_2(x',x_2)h_1(x_2,x). \tag{1}$$

For simplicity, but without loss of generality, we have assumed a one-dimensional geometry.

The source is a thin nonlinear crystal from which photons are emitted in pairs, signal and idler, by a process of spontaneous parametric downconverion in the presence of a pump. We assume that the signal and idler have the same wavelength and travel through the same optical system. Since the photon pairs are emitted independently from different positions of the crystal, the second-order coherence function at the aperture is related to the rate of emission at the source by [1,6]

$$G_A^{(1)}(x_1,x_2) \propto \int I_p(x) h_1^*(x_1,x) h_1(x_2,x) dx, \tag{2}$$

where $I_p(x) = |E_p(x)|^2$ and $E_p(x)$ is the pump field. This function is separable for a narrow pump and becomes less separable as the pump size increases. Full separability corresponds to complete coherence [1]. The detector measures the intensity (the photon detection rate) at the detection plane. This is related to the coherence function at the slits by

$$I(x') = \left[|h_2(x',x_1)|^2 G_A^{(1)}(x_1,x_1) + |h_2(x',x_2)|^2 G_A^{(1)}(x_2,x_2)\right] + \left[h_2^*(x',x_1)h_2(x',x_2)G_A^{(1)}(x_1,x_2) + c.c.\right].$$

(3)

The third and fourth terms of Eq. (3), which constitute the interference terms, are proportional to the coherence function at the two slits.

The fourth-order coherence properties may be determined from the two-photon wavefunction at the slits, which is related to the pump field $E_p(x)$ by [6,13]

$$\Psi_A(x_1,x_2) \propto \int E_p(x) h_1(x_1,x) h_1(x_2,x) dx.$$  (4)

This expression is mathematically similar to Eq. (2), with the pump field playing the role of the source intensity. The detector measures the rate of coincidence of photons at pairs of points in the detection plane,

$$G^{(2)}(x',x'') = |\Psi(x',x'')|^2$$  (5)

where

$$\Psi(x',x'') = h_2(x',x_1)h_2(x'',x_1)\Psi_A(x_1,x_1) + h_2(x',x_2)h_2(x'',x_2)\Psi_A(x_2,x_2)$$
$$[h_2(x',x_1)h_2(x'',x_2) + h_2(x',x_2)h_2(x'',x_1)]\Psi_A(x_1,x_2),$$  (6)

Equation (6) may be obtained by using the relation $\Psi(x',x'') \propto \int E_p(x) h(x',x) h(x'',x) dx$ together with Eqs. (1) and (4). The third and fourth terms of (6), the interference terms, are proportional to the two-photon wavefunction at the slits. Again, the similarity between Eq. (6) and Eq. (3) is notable.

We now consider a specific case for which the optical system between the slits and the detector is a Fourier-transform system (a lens in a 2-$f$ configuration), i.e., $h_2(x,x_1) \propto \exp(-i\frac{2\pi}{\lambda f} x x_1)$, where $f$ is the focal length and $\lambda$ is the wavelength of the signal/idler. For simplicity, we assume that the slits are located symmetrically above and below the optical axis, say at $x_1 = -a/2$, and $x_2 = a/2$, and that the source is symmetric, so that $G_A^{(1)}(x_1,x_1) = G_A^{(1)}(x_2,x_2)$, $\Psi_A(x_1,x_1) = \Psi_A(x_2,x_2)$, and $G_A^{(1)}(x_1,x_2)$ and $\Psi_A(x_1,x_2)$ are real functions. Under these conditions, Eq. (3) gives

$$I(x') \propto 1 + g_A^{(1)} \cos\left(2\pi \frac{x'}{\Lambda}\right), \tag{7}$$

where $g_A^{(1)} = \frac{G_A^{(1)}(\frac{a}{2},-\frac{a}{2})}{G_A^{(1)}(\frac{a}{2},\frac{a}{2},)}$ is the degree of coherence at the slits and $\Lambda = \frac{\lambda f}{a}$. This is a fringe pattern with period $\Lambda$ and visibility

$$V_1 = g_A^{(1)}, \tag{8}$$

equal to the degree of coherence at the slits.

Likewise, Eq. (6) leads to

$$G^{(2)}(x',x'') \propto \left|\cos\left(\pi\frac{x'+x''}{\Lambda}\right) + \psi_A \cos\left(\pi\frac{x'-x''}{\Lambda}\right)\right|^2$$
$$= 1 + \frac{1}{1+\psi_A^2}\cos\left(2\pi\frac{x'+x''}{\Lambda}\right) + \frac{\psi_A^2}{1+\psi_A^2}\cos\left(2\pi\frac{x'-x''}{\Lambda}\right) + \frac{2\psi_A}{1+\psi_A^2}\left[\cos\left(2\pi\frac{x'}{\Lambda}\right) + \cos\left(2\pi\frac{x''}{\Lambda}\right)\right],$$
$$\tag{9}$$

where $\psi_A = \frac{\Psi_A(\frac{a}{2},-\frac{a}{2})}{\Psi_A(\frac{a}{2},\frac{a}{2},)}$, and $G^{(2)}(x',x'')$ is normalized so that its integral is unity, i.e., $G^{(2)}(x',x'')$ represents the joint probability density of detecting a photon at $x'$ and another at $x''$. This 2-D fringe pattern is a result of a combination of single-photon and two-photon interference effects; the former is obtained by integrating $G^{(2)}(x',x'')$ with respect to $x''$ to obtain the marginal probability density of detecting a single-photon at $x'$ given that the other is detected anywhere:

$$I_m(x') \propto 1 + V_{1m} \cos\left(2\pi \frac{x'}{\Lambda}\right), \tag{10}$$

where

$$V_{1m} = \frac{2\psi_A}{1 + \psi_A^2}. \tag{11}$$

This is a sinusoidal pattern of visibility $V_{1m}$.

To determine the pure two-photon interference we define the excess coherence function,

$$\Delta G^{(2)}(x',x'') = G^{(2)}(x',x'') - I_m(x')I_m(x'') + A \tag{12}$$

by subtracting the product of the marginal rates and adding a constant $A$ to account for duplicate subtraction of a background term [3, 5]. The parameter $A$ is calculated by normalizing the excess coherence function such that it integrates to unity over the region of interest. It follows from Eqs. (9) and (10) that

$$\begin{aligned}\Delta G^{(2)}(x',x'') &= V_{12} \sin\left(2\pi \frac{x'}{\Lambda}\right)\sin\left(2\pi \frac{x''}{\Lambda}\right) + V_{12}^2 \cos\left(2\pi \frac{x'}{\Lambda}\right)\cos\left(2\pi \frac{x''}{\Lambda}\right) + A \\ &= V_{12}(1+V_{12})\cos\left(2\pi \frac{x'+x''}{\Lambda}\right) + V_{12}(1-V_{12})\cos\left(2\pi \frac{x'-x''}{\Lambda}\right) + A,\end{aligned} \tag{13}$$

where

$$V_{12} = \frac{1 - \psi_A^2}{1 + \psi_A^2}. \tag{14}$$

This 2-D pattern has visibility $V_{12}$. It follows from Eqs. (11) and (14) that

$$V_{12}^2 + V_{1m}^2 = 1. \tag{15a}$$

Equation (15a) is a complementarity relation between the visibilities of the single- and two-photon interference for light generated by a thin two-photon light source [2-5]. It has been shown [6] that a similar relation applies for a thick crystal. Both single- and two-photon visibilities are determined by the normalized two-photon wavefunction $\psi_A$ at the slits. If the illumination optical system is also a 2-$f$ system, i.e., $h_1(x,x_1) \propto \exp(-i\frac{2\pi}{\lambda f} x x_1)$, then $\psi_A$ is proportional to the Fourier transform of the pump distribution, $\psi_A = \tilde{E}_p\left(\frac{2\pi}{\Lambda}\right) \Big/ \tilde{E}_p(0)$, where the tilda indicates the Fourier transform operation. If the pump is a unifrom beam of width $b$, then $\psi_A = \mathrm{sinc}(b/\Lambda) = \mathrm{sinc}(ba/\lambda f)$, i.e., $\psi_A$ is governed by the angular size of the source.

The relation between the "pure" and "marginal" single-photon visibilities, $V_1$ and $V_{1m}$, respectively, may be established by determining the relation between the degree of coherence $g_A^{(1)}$ and the normalized two-photon wavefunction $\psi_A$. In view of Eqs. (2) and (4), these numbers are related. If the pump field is a real rectangular function, then $g_A^{(1)} = \psi_A$, so that $V_1 = \psi_A$ and $V_{1m} = 2V_1/(1+V_1^2)$ is a monotonic increasing function of $V_1$. It also follows that

$$V_{12} = \frac{1-V_1^2}{1+V_1^2}. \tag{15b}$$

This monotonic decreasing function establishes another complementarity relation similar to that in Eq. (15a), as illustrated in Fig. 2.

The analogy between the entanglement properties of light emitted from a SPDC source and the properties of light emitted from an incoherent source can be extended to include temporal/spectral effects. The analogy is also applicable to thick sources, but the nature of equivalence is somewhat different [6]. These more general results are not germane to the complementarity relations derived above; the reader is referred to Ref. [6] for further details.

## III. EXPERIMENT

The experimental setup is illustrated in Fig. 3. A 1-mm thick $LiO_3$ crystal of 6x6 mm cross-section is pumped by a 35-mW $Kr^+$-ion laser of 406-nm wavelength to generate spontaneous parametric down-converted light in the degenerate collinear Type-I configuration. The down-converted beam is passed through a circular aperture of 2-mm diameter and through a double-slit aperture at a variable distance $d$ from the circular aperture. The slits are of width 0.35 mm each and are separated by a distance $a = 0.70$ mm. The unconverted pump is prevented from reaching the slits by use of a combination of two Glan Thompson prisms oriented with orthogonal polarizations before and after the crystal. The double-slit aperture is followed by a 2-$f$ system using a C-mount Nikon lens of focal length $f = 50$ mm. The detector is an intensified CCD camera (ANDOR ICCD-432, model DH5H7-18F-31, with a VIS, 420-920 nm, photocathode) with 512x512pixels, each of size 24x24 $\mu m^2$. The camera is cooled to $-25^{o}C$ and has quantum efficiency $\eta=0.5$ at a wavelength of 812 nm. The digitized analog signal is transferred to a computer for subsequent analysis.

Conventional interference patterns were measured by recording the average image detected by the camera over an exposure time of 2 seconds. Each coincidence interference pattern, however, was obtained by a procedure based on a sequence of 240,000 frames collected in a total of 48 seconds. Each frame is thresholded, which results in an array of 0's and 1's. Since the detection of a photon is typically marked by a patch of 1's extending over a neighborhood of 3x3 pixels, we identify the locations of registered photons by locating such 3x3 patches. Within each patch, the pixel with the highest analog signal marks the photon location.

Most of the frames are empty (all 0's). Frames with two registered photons (two 1's) are the useful ones, which we use to determine the coincidence rate function, and other frames are disregarded.

Since the down-converted beam has circular symmetry whereas the slits do not, the system is inherently two-dimensional. In order to avoid the complexity of determining coincidence rates at all pairs of points, i.e., dealing with four-dimensional data, we have limited ourselves to collecting data from a narrow rectangular strip across the CCD in the middle of the observed pattern. Also, we only consider frames featuring pairs for which the vertical separation of the photon registrations is less than one-third of the horizontal separation. Additionally, the data obtained from each frame are reduced to a one dimensional vector $\mathbf{X}$ with all 0's and only two 1's. The average of the matrices $\mathbf{X}^T\mathbf{X}$ (where $^T$ indicates transpose) for all frames provides an estimate of the function $G^{(2)}(x_1,x_2)$ at the positions of the pixels. This procedure offers an estimate of $G^{(2)}(x_1,x_2)$ with a spatial resolution limited to 3 pixels. It cannot provide an estimate for $G^{(2)}(x,x)$ since it can only register spatially separated photons. This value may be interpolated from neighboring points. The accuracy of this data processing technique was verified by integrating $G^{(2)}(x_1,x_2)$ with respect to $x_1$ or $x_2$ to determine the single-photon rate. The result is approximately the same as a cross section of the diffraction pattern.

The experiment was repeated 5 times at different angular sizes of the source, ranging from near field to far field ($d$ = 5.5, 6.3, 30, 54, and 87 cm). In each case the one- and two-photon interference patterns were determined and the corresponding visibilities were estimated. Samples of these patterns are shown in Figs. 4-6. In these figures, all pixels are superpixels of size 4x4 pixels of the CCD chip.

Figure 4 shows the single-photon interference pattern in the far field, $d = 54$ cm, and the near field, $d = 5.5$ cm. These are simply conventional diffraction patterns for a double-slit aperture, accumulated in a 2 sec. exposure. The interference fringes are clearly visible in the far field case, and are not visible in the near field case. This is of course expected since the degree of coherence is diminished in the near field case.

Examples of the measured two-photon coincidence rate $G^{(2)}(x',x'')$ and excess coincidence rate $\Delta G^{(2)}(x',x'')$ are shown in Figs. 5 and 6. Figure 5 shows $G^{(2)}(x',x'')$ in the far field, $d = 87$ cm. This function is nearly separable (unentangled) and the corresponding excess coincidence rate $\Delta G^{(2)}(x',x'')$ is approximately flat. Figure 6 shows $G^{(2)}(x',x'')$ and $\Delta G^{(2)}(x',x'')$ in the near case, $d = 6.3$ cm. Here, the function $\Delta G^{(2)}(x',x'')$ clearly exhibits modulation along the direction $x' + x''$ =constant, which is indicative of the lack of separability.

The visibilities of single- and two-photon interference, $V_{1m}$ and $V_{12}$ respectively, for the five experiments are displayed in Fig. 7, together with a plot of the theoretical complementarity relation (solid curve), $V_{12}^2 + V_{1m}^2 = 1$, as given in Eq. (15a). The ideal relationship is derived under the assumption of a thin crystal and narrow spectral SPDC bandwidth.

The dashed curve in Fig. 7, which was obtained by simulating Eqs. (5.6, 5.8, and 5.10) in Ref. [6] and using an optical-system impulse response function representing free-space propagation from the crystal to the double slits and the ensuing 2-$f$ system. This accommodates a crystal of finite thickness; however, the down-converted light was taken to be narrow with respect to the central down-conversion wavelength. The visibilities $V_{1m}$ and $V_{12}$ were calculated for various values of the distance $d$ and the dashed curve is a fit to these calculated points. Note that the simulated complementarity curve (dashed) lies below the ideal complementarity curve (solid) demonstrating that taking the thickness of the crystal into consideration lowers the resulting curve. The experimental points include the effect of finite bandwidth and are, therefore, expected to be even lower, as they indeed are.

## IV. CONCLUSION

We have measured the visibilities of second-order (one-photon) and fourth-order (two-photon) interference fringes in a Young's double-slit experiment carried out with light generated by spontaneous parametric down-conversion. We conducted these experiments using an intensified CCD (ICCD) camera, which records the photon arrival at all spatial points within the same time window, thus overcoming a measurement loophole associated with the more-common method of using scanning point detectors, as in previous photon-coincidence measurements. The use of an ICCD camera to measure photon coincidences was first suggested by Klyshko [13] and its experimental use was first reported by Jost *et al*. [14].

As the effective source size is increased, the visibility of single-photon interference decreases while the visibility of two-photon interference increases. This is the first experimental demonstration of the complementarity between single- and two-photon interference in the spatial domain. The origin of this complementarity is the opposite roles played by separability on coherence and entanglement. As we move from the far field to the near field, the following sequence takes place: the effective source size increases, the separability decreases, the coherence decreases, the visibility of single-photon interference decreases, the entanglement increases, and the visibility of two-photon interference increases.

## ACKNOWLEDGMENTS

We are grateful to Abner Shimony and Michael Horne for valuable and illuminating discussions regarding the theoretical aspects of this problem. This research was supported by the National Science Foundation.

**Figure captions**

Fig. 1 A generic Young's double-slit optical system.

Fig. 2 Complementarity relations: pure and marginal single-photon visibilities, $V_1$ and $V_{1m}$, respectively, versus two-photon visibility $V_{12}$.

Fig. 3. Experimental arrangement.

Fig. 4 Single-photon interference pattern at: a) Far field ($d$ = 54 cm), and b) Near field ($d$ = 5.5 cm). The grey scale is normalized as indicated in both a) and b).

Fig. 5. Two-photon coincidence rate $G^{(2)}(x',x'')$ in the far field ($d$ = 870 mm).

Fig. 6. Two-photon interference in the near field ($d$ = 6.3 cm): a) coincidence-rate $G^{(2)}(x',x'')$, b) excess coincidence rate $\Delta G^{(2)}(x',x'')$.

Fig. 7. The marginal one-photon visibility $V_{1m}$ versus the two-photon visibility $V_{12}$. Solid curve represents the ideal complementarity relationship $V_{12}^2 + V_{1m}^2 = 1$. The experimental results are indicated by circles. Dashed curve represents the results of a simulation that takes into consideration the experimental geometry and the crystal thickness.

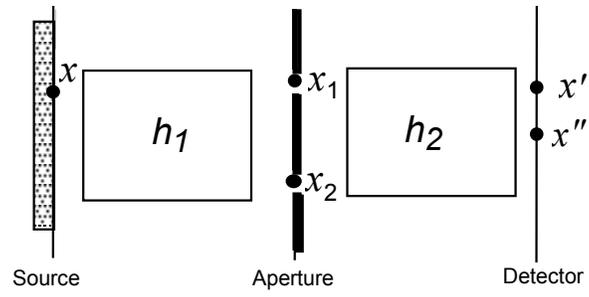

Fig. 1 A generic Young's double-slit optical system.

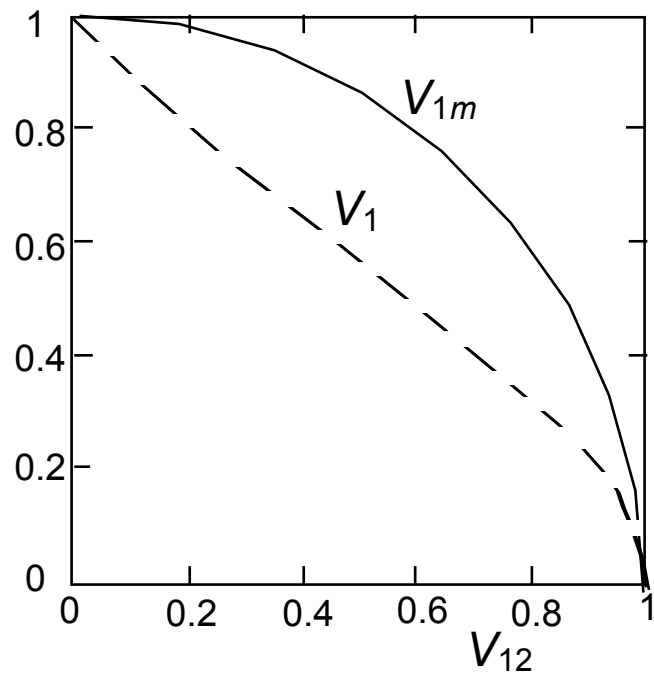

Fig. 2 Complementarity relations: pure and marginal single-photon visibilities, $V_1$ and $V_{1m}$, respectively, versus two-photon visibility $V_{12}$.

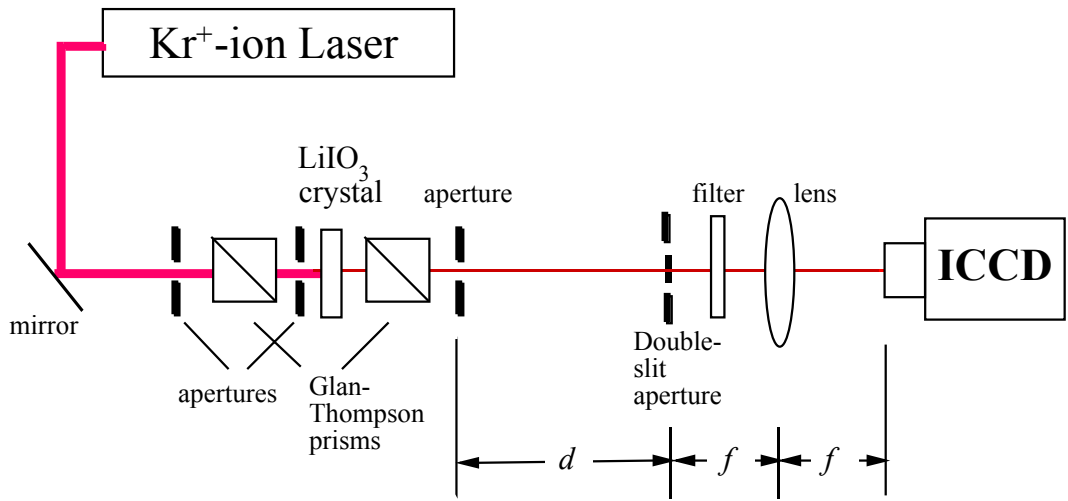

Fig. 3.  Experimental arrangement.

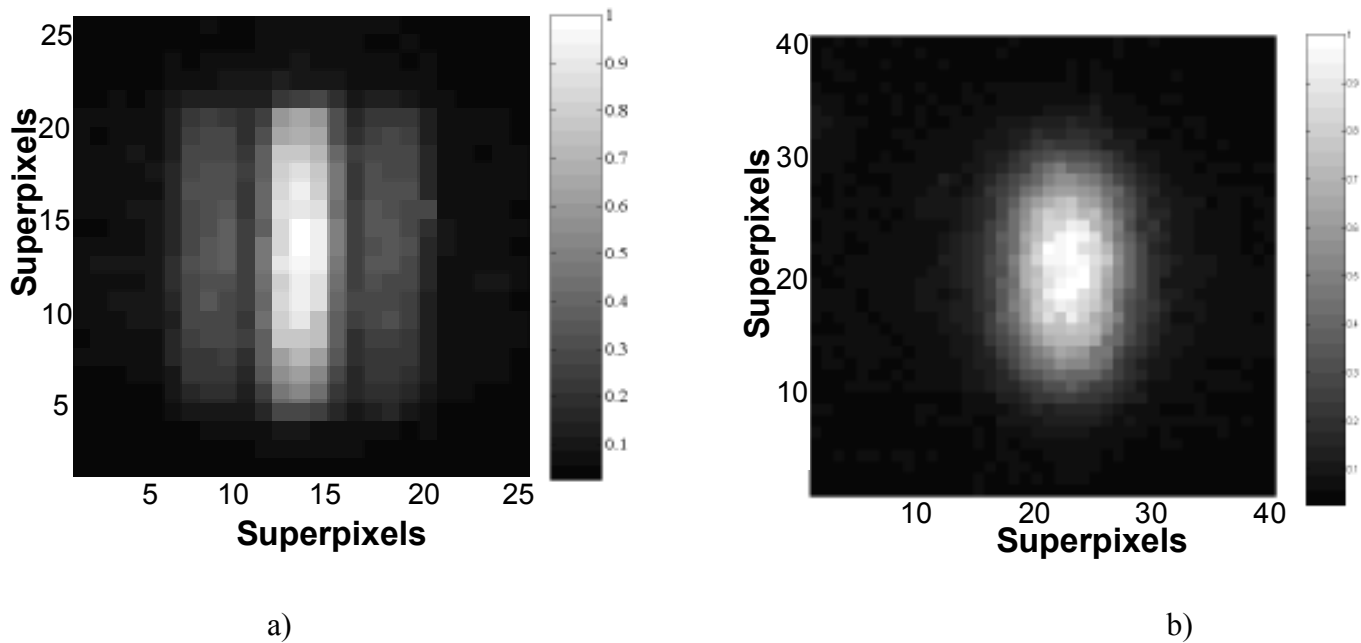

a)

b)

Fig. 4  Single-photon interference pattern at:  a) Far field ($d = 54$ cm), and b) Near field ($d = 5.5$ cm). The grey scale is normalized as indicated in both a) and b).

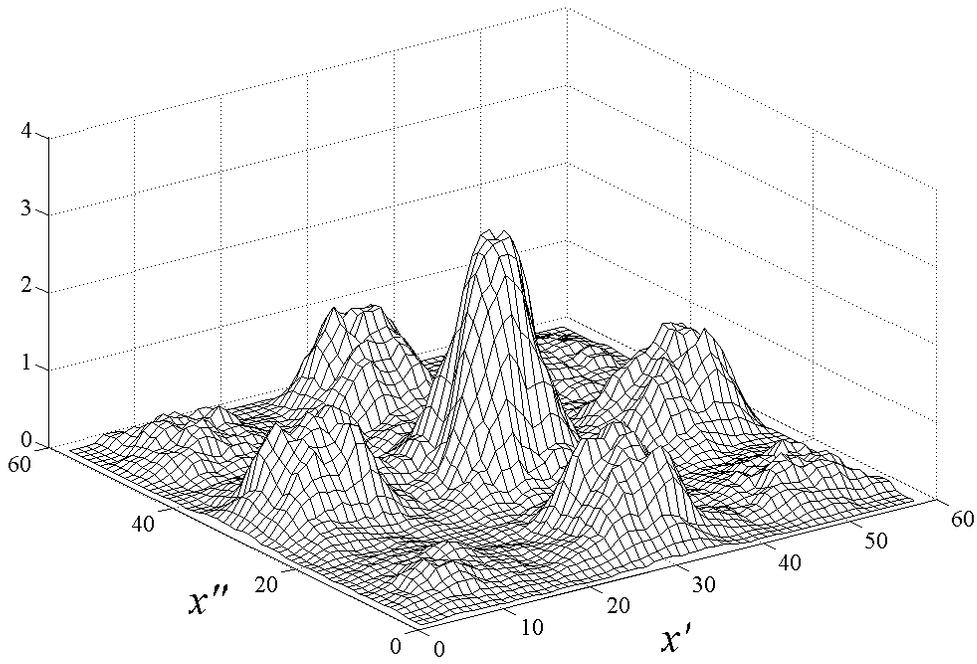

Fig. 5. Two-photon coincidence rate $G^{(2)}(x', x'')$ in the far field ($d$ = 870 mm).

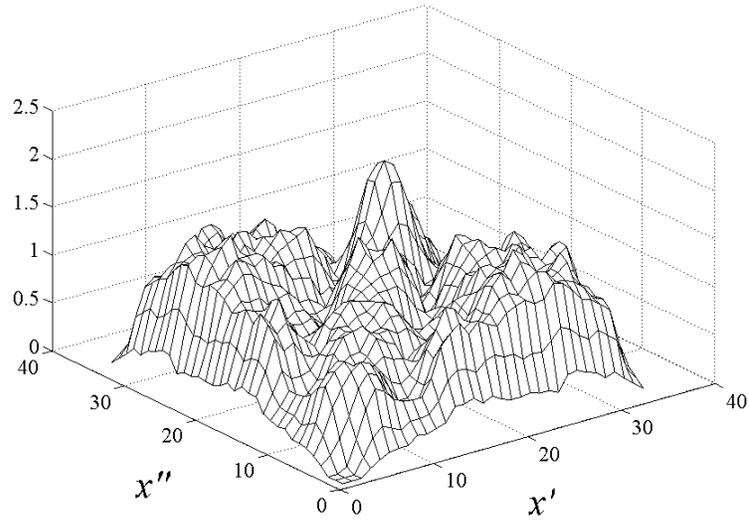

(a)

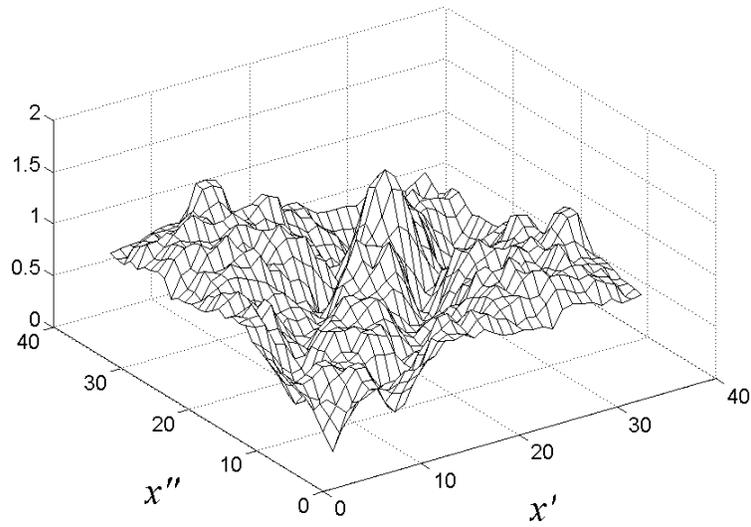

(b)

Fig. 6. Two-photon interference in the near field ($d = 6.3$ cm): a) coincidence-rate $G^{(2)}(x', x'')$, b) excess coincidence rate $\Delta G^{(2)}(x', x'')$.

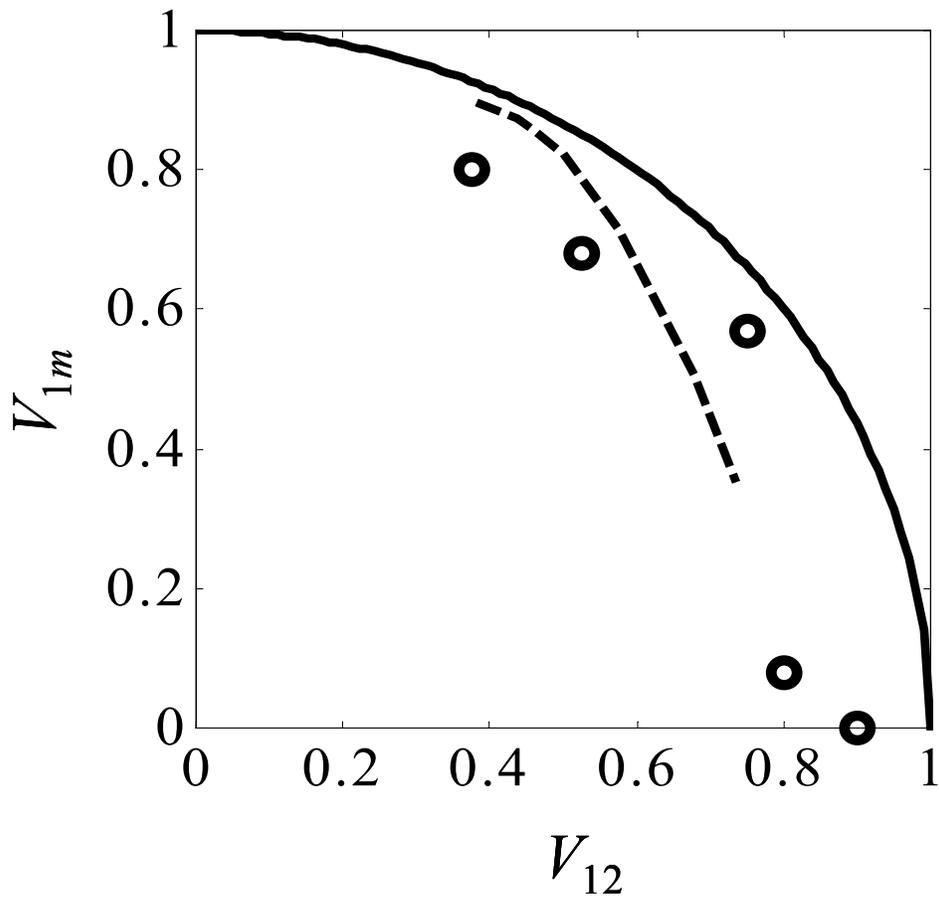

Fig. 7. The marginal one-photon visibility $V_{1m}$ versus the two-photon visibility $V_{12}$. Solid curve represents the ideal complementarity relationship $V_{12}^2 + V_{1m}^2 = 1$. The experimental results are indicated by circles. Dashed curve represents the results of a simulation that takes into consideration the experimental geometry and the crystal thickness.